\documentclass[aps,prb,showkeys,showpacs,superscriptaddress,twocolumn,floatfix]{revtex4}
\usepackage{amsmath}
\usepackage{amsfonts}
\usepackage{graphicx}
\usepackage{psfrag}
\usepackage{hyperref}

\begin{document}

\title{Norm breaking in a queue---athermal phase transition}

\author{Krzysztof Malarz}
\homepage{http://home.agh.edu.pl/malarz/}
\email{malarz@agh.edu.pl}
\affiliation{
AGH University of Science and Technology,
Faculty of Physics and Applied Computer Science,\\
al. Mickiewicza 30, PL-30059 Krak\'ow, Poland
}

\author{Ruediger Korff}
\affiliation{
Mainland Southeast Asian Studies, Universit\"at Passau,
Innstrasse 43, D-94032 Passau, Germany
}

\author{Krzysztof Ku{\l}akowski}
\email{kulakowski@novell.ftj.agh.edu.pl}
\affiliation{
AGH University of Science and Technology,
Faculty of Physics and Applied Computer Science,\\
al. Mickiewicza 30, PL-30059 Krak\'ow, Poland
}

\date{\today}
 
\begin{abstract}
Standard order-disorder phase transition in the Ising model is described in terms of rates of processes of spin flips. This formulation 
allows to extend numerous results on phase transition for sciences other than physics of magnetism. We apply it to the problem of 
norm breaking. Two strategies: to cooperate or to defect, form an analogy to two spin orientations. An example is a crowd at an exit, where 
to defect means to push others, and to cooperate is to move slowly together. 
\end{abstract}
 


\maketitle

%% ############################################################################
\section{Introduction}
%% ############################################################################

Phase transitions are ubiquitous in nature. The ferro-paramagnetic phase transition in Ising system\cite{huang} is an example cited perhaps most frequently. The concept of phase transition is valid and useful also in social sciences, but the example taken from magnetism seems to dominate minds.
As an effect, we read on social temperature\cite{soctem1,soctem2,soctem3,soctem4} and social energy\cite{socer}, although these physical quantities have no direct counterparts in sociology. Here we argue that in social and presumably also other sciences, it is more natural to describe a phase transition in terms of rates of processes, which appear in the problem under consideration. Further, the standard magnetic phase transition can be also described in terms of rates of processes: in the simplest one-spin mean field model, there are two processes: the spin flip from up to down and from down to up. This indicates, that the approach advocated here is more general. A similar scheme has been proposed by Galam and coworkers on iterative equations, with discrete time\cite{ga1,ga2}. Here we work with Monte Carlo simulations, supplemented with differential Master Equations. 

We do not claim that concepts of thermodynamics is not applicable in social sciences. In particular, consider a large ensemble of similar and independent events, each described with a variable $x$. Analysis of the probability distribution of $x$, when combined with the maximum entropy principle, can give a valuable insight into the nature of underlying processes. A recent example of this procedure can be found in Ref. \onlinecite{ara}. To accomplish this, however, we need an assumption on the equilibrium state of the system, what is rarely firm in sociology and can only be justified ex post, from pertinence of obtained results. On the other hand, rates of changes of investigated variables allow to reproduce time dependences of investigated variables, and not only their equilibrium state.
 
As an example, we chose the problem of norm breaking at a crowded exit. Crowd simulation become an independent branch of computational science, with 
numerous applications\cite{cr1,cr2,cr3}. For our purposes, the system can be reduced to a set of cells occupied by pedestrians. We assume that the crowd is temporarily jammed, therefore pedestrians cannot move. However, each pedestrian can push other pedestrian before him, to activate the motion as soon as possible and to reach the exit, or at least to force people before her/him to push as well. This small psychological advantage, however, can have fatal consequences, as the force exerted by particular pedestrians accumulates at the exit and people there can be badly injured, if the crowd is large and the selfish pushing strategy prevails. This aspects allows to interpret the situation as an example of the Prisonner's Dilemma\cite{phd}. The connection between crowd dynamics and game theory has been emphasized by Dirk Helbing\cite{cvss}. In particular, we expect that the phase transition in minds of pedestrians, which we investigate here, is coupled to the phase transitions discussed in theory of crowd dynamics, where the stop-and-go phase and crowd turbulence have been distinguished\cite{dh1,dh2}.  

The text is constructed as follows. In the next section, the Ising phase transition is described in terms of rates of spin-flips. We provide an example, where the rates are taken from the standard thermodynamic, temperature-dependent formulation. For this case computer simulation reproduces the exact Onsager solution\cite{Onsager}. We propose also a mean field formulation of the problem in terms of Master Equations. As a rule, the mean field approach gives a value of the Curie temperature, which is larger than the approximately true value obtained in the simulation. In the thermodynamic case, the mean-field solution is obtained which is slightly better than the Bragg--Williams solution but worse than the Bethe--Peierls solution\cite{huang}. Here, the criterion is the distance between the obtained critical point and its exact value. Section \ref{sec3} is devoted to the crowd problem, as remarked above. As the full formulation of the problem on the square lattice contains many parameters, analytical calculations are presented for two one-dimensional cases, where {\it i)} two strategies are symmetric, {\it ii)} the homogeneous states are absorbing. Whilst the fork bifurcation\cite{glen} is found in the case {\it i)}, in the case {\it ii)} the bifurcation is transcritical. Numerical calculations are presented also for pedestrians in cells of a square lattice, with model set of rates. In the same section we demonstrate that the relaxation time $\tau$ goes to infinity at the critical subspace of the spin-flip rates. Last section \ref{sec4} is devoted to conclusions.

%% ############################################################################
\section{Athermal Ising phase transition\label{sec2}}
%% ############################################################################

In physics, the ferro-paramagnetic transition of Ising spins $s=\pm 1$ at the square lattice is referred to most often. In this transition, the energy of the exchange interaction between nearest neighbours competes with the thermal noise. As a result, the ferromagnetic phase appears at low temperature, whilst at high temperature the directions of spins are disordered. The same content can be expressed as follows. Spins flip from up to down and the oppositely, but the rates of these flippings depend on the states of spins in their direct neighborhood. In square lattice a spin up surrounded by four spins up flips rather rarely. On the other hand, a spin up surrounded by four spins down flips likely. Let us denote these rates as $w_n$, where $n=0,\cdots,4$ marks the number of spins at nearest four nodes in the same state as the spin under consideration. Then, a spin flips from up to down with rate $w_n$ if it is surrounded by $n$ spins up and $(4-n)$ spins down. From the up-down symmetry, a spin down surrounded by $n$ spins down flips to the up orientation with the same rate $w_n$.

\begin{figure}[ht]
\psfrag{m}{$m$}
\psfrag{x}{$x$}
\includegraphics[width=.49\textwidth]{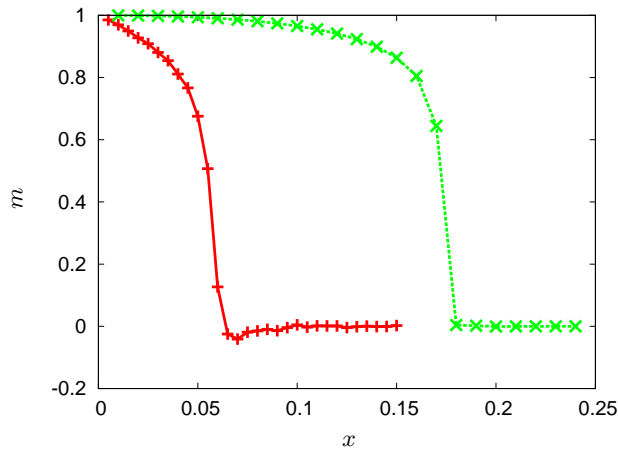}
\caption{Mean value $m$ of the variable $s=\pm 1$ (points) against parameter $x$ in a square lattice. When the rates are selected as $w_0=w_1=1$, $w_2=1/2$, $w_3=x$ and $w_4=x^2$, the system dynamics is equivalent to the Metropolis algorithm, where $x=\exp(-4\beta J)$. We obtain the curve of crosses ($\times$) and the transition at $x=0.17$. For the rates $w_0=1$, $w_1=2x$, $w_2=x$, $w_3=x/2$ and $w_4=x/4$ we get the curve of pluses ($+$), where the ordered phase ends at about $x=0.06$. The simulations were carried out for square lattice with $N=100^2$ spins and helical boundary conditions. The spins were visited in type-writer order. For given $x$ parameter $T_{\text{max}}=10^7$ Monte Carlo steps per spin were performed. The mean value of magnetization $m(x)$ was evaluated as time average over the last $T_{\text{max}}/2=5\cdot 10^6$ steps.}
\label{fig-1}
\end{figure}

In statistical mechanics, spins flip with the rates which fulfill the condition of detailed balance. For a spin up in the square lattice with $n$ neighbours up, this condition is $aw_n=bw_{4-n}$, where $a=(1+m)/2$ is the probability of spin up, $b=(1-m)/2$ is the probability of spin down, and $m=\sum_i s_i/N$, where summation is over $N$ sites $i$. As the ratio $a/b=\gamma^{4(n-2)}$ in equilibrium (where $\gamma=\exp(\beta J)$, $\beta=1/T$ is the inverse temperature and $J$ is the exchange integral), and the rates $w_n$ do not depend on the fact if the system is in equilibrium or not, these rates should also fulfill the condition $w_{4-n}/w_n=\gamma^{4(n-2)}$. This is fulfilled in particular for the Metropolis algorithm\cite{Metropolis}, where 
$w_0=w_1=1$, $w_2=1/2$, $w_3=\exp(-4\beta J)$ and $w_4=w_3^2$. When these rates are introduced to the Monte Carlo code, the obtained magnetization vanishes at $w_3=0.17$, as shown in Fig. \ref{fig-1}. This gives the critical temperature $T_C/J=2.27$ , what agrees well with Onsager solution\cite{Onsager}.

The conditions of detailed balance fix the relations between $w_n$ and $w_{z-n}$ only, where $z$ is the number of nearest neighbours. Then the relation between, say, $w_0$ and $w_1$ remains unspecified. In particular, we can set $w_n=\gamma ^{4-2n}$ However, in numerical calculations we are interested in high rates, to speed up the calculations, and therefore the Metropolis version is more convenient.  In our text, the Metropolis rates are used for the sake of comparison of our results with the special case, where the detailed balance conditions are fulfilled. In general, the thermodynamic balance conditions are not valid, and our rates do not fulfill them. 

\begin{widetext}
The related mean field theory can be formulated in terms of the Master Equations\cite{gard}
\begin{equation}
\frac{da}{dt}=\sum_{n=0}^z \frac{z!}{n!(z-n)!}a^nb^{z-n}(-aw_n+bw_{z-n}).
\label{eq1}
\end{equation}
To check its result again in the thermodynamic case, we can put $w_n=\gamma ^{4-2n}$. Substituting this to Eq. \eqref{eq1}, after simple manipulations we get
\begin{equation}
\label{eq2}
\frac{da}{dt}=-a(a/\gamma+b\gamma)^z+b(a\gamma+b/\gamma)^z.
\end{equation}
In this symmetric case, the stationary solution $a=b$ $(m=0)$ always exists, but it ceases stability when 
\begin{equation}
\label{eq3}
\frac{d}{dm}\frac{da(m)}{dt}=0
\end{equation}
at $m=0$, what gives $\tanh(\beta J)=1/z$ at $T_C$. This result is slightly better than the simplest Bragg--Williams mean field, where no correlations exist and $z\beta J=1$. However, it is worse than the Bethe--Peierls result that $\tanh(\beta J)=1/(z-1)$ at $T_C$. This is because the states of next neighbours are not correlated with each other; in Eq. \eqref{eq1}, their weights are given by the Bernoulli distribution. For $z=4$, the stationary solution for the ordered phase is
\begin{equation}
\label{eq4}
m=\frac{\sqrt{-5+4y+6y^2-4y^3-y^4+2\sqrt{(y-1)^4(y+1)^2(5+2y+y^2)}}}{(y-1)^2},
\end{equation}
where $y=\gamma ^2=\exp(2\beta J)$.
\end{widetext}

We could also try to use the `Metropolis rates' $w_0=w_1=1$, $w_2=1/2$, $w_3=x$ and $w_4=x^2$, to construct the Master Equation. In this case, the critical value of $x$ is about 0.572. Substituting $x=\exp(-4\beta J)$, we get $T_C=7.14 J$. This is much worse even than the Bragg--Williams solution $T_C=4$, while the rates $w_n=\gamma ^{4-2n}$ give $T_C=3.91$, the Bethe--Peierls solution is $T_C=2.89$ and the exact solution is $T_C=2.27$ in units of $J$. This result is not strange, because in fact the detailed balance conditions are not fulfilled in the stationary state obtained within the mean field theory. Instead, the contributing processes on the right side on Eq. \eqref{eq1} mutually cancel at the stationary state. When we put $w_1=1$, $w_2=1/2$ and $w_3=x$, the processes of flipping of spins with one, two or three neighbours in the same state are much faster, than they should be if $w_0=1$, and the balance of processes is disturbed.

The time $\tau$ of relaxation to a stable state can be evaluated by linearization procedure. As long as the state $m=0$ is stable, the equation of motion can be written for small $m$ as 
\begin{equation}
\label{eq5}
\frac{da(m)}{dt}\approx \lambda m,
\end{equation}
where 
\begin{equation}
\label{eq6}
\lambda=\frac{d}{dm}\frac{da(m)}{dt}
\end{equation}
at $m=0$, and $\lambda<0$. The approximate solution is $m(t)=m(0)\exp(\lambda t)$. As $\tau$ is usually defined by the relation $m(t)=m(0)\exp(-t/\tau)$, we have $\tau=-1/\lambda$.

To conclude this section, Eq. \eqref{eq1} provides the generalization of the mean field theory for non-physical (non-thermodynamic) cases. This means that the rates $w_n$ do not fulfil the thermodynamic conditions of detailed balance, but they are mutually independent. Also, their use does not imply any form of energy or temperature. As we have shown, phase transitions appear as a consequence of interplay of flows of probability between different states.

%% ############################################################################
\section{Norm breaking in crowd\label{sec3}}
%% ############################################################################

There is a countless number of theories of human behaviour, each one bearing some element of reductionism. Our choice is the Scheff theory of pride and shame\cite{scheff,tursim}. There are two loops of behaviour, the shame loop and the pride loop, which relate to bad and good self-evaluation\cite{tursim}. Accordingly, agents in the pride loop are more willing to select altruistic strategies, while those in the shame loop are likely to select the selfish one. Additionally, in the pride loop the communication with other agents is more efficient. Being in a loop, an agent prefers to remain there, hence the term; still, there is some probability to switch. In particular, a friendly communication with other agents can trigger the switch from the shame loop to the pride loop.

When we project this theory to the problem of norm breaking, and to our example of the crowd behaviour in particular, it is clear that both the selfish strategy and the cooperative strategy have the property of self-stabilization. Further, the selfish strategy can spread; being pushed by other agent reduces the pride and drives the self-evaluation down. On the other hand, an example of a calm behaviour can bear an attitude to copy this strategy as well. This mechanism works better, if more neighbours behave in the same way\cite{threshold,sznajd}. Concluding, it is justified to introduce the rates of strategy changes, which depend on the number of neighbours which play a given strategy.

In principle, the process can be investigated in any space dimension, but the number of parameters $w_n$ increases with the number $z$ of nearest neighbours. To keep the analytical description simple, let us consider a one-dimensional queue, where a pedestrian has only two neighbours. As it is known, the ferro-paramagnetic phase transition in the one-dimensional Ising chain appears only at T=0 K. Here its presence is a consequence of the mean field approach. Now two orientations of spin mean two options: the selfish strategy with probability $a$ and the cooperative strategy with probability $b=1-a$. In a linear queue, there are two neighbours only. In the `additive' case, i.e. if neighbours before and behind a given pedestrian are equivalent, the equation can be written as
\begin{equation}
\label{eq7}
\frac{da}{dt}=a^2(-aw_2+bw_0)+2ab(-aw_1+bw_1)+b^2(-aw_0+bw_2),
\end{equation}
where three consecutive terms on r.h.s. refer to three configurations of the neighborhood: two selfish neighbours ($a^2$), one selfish and one quiet ($2ab$) and two quiet ones ($b^2$). As we have described the change of each strategy with the same parameters, again the solution $a=b$ is always possible. As before, the critical point can be identified from the instability of this solution. Denoting the ratio $(w_0-2w_1)/w_2$ as $\phi$, we get the solution for $m$ different from zero and stable if $\phi>3$. There
\begin{equation}
\label{eq8}
m^2=\frac{\phi-3}{\phi+1},
\end{equation}
otherwise $m=0$. The counterpart of the critical temperature is then the value $\phi=3$. This is an example of the fork bifurcation. To have an ordered state, where most pedestrians apply the selfish strategy or most pedestrians apply the quiet (cooperative) strategy, we have to keep the rate $w_0$ large enough. This means that a pedestrian should change its state quickly enough when it is different than the state of both her/his neighbours.

In our case, we get the relaxation time 
\begin{equation}
\label{eq9}
\tau=\frac{2}{w_2(3-\phi)}
\end{equation}
and $\tau$ tends to infinity as $\phi$ tends to its critical value $\phi=3$ from below. This means, that the mean-field version of `critical slowing down' is present here.

In the case of queue, we would like to consider also the case when a pedestrian makes a difference between the states $ab$ and $ba$; it is perhaps not the same situation to be pushed from behind or just to observe that the neighbour in front of you is pushing somebody else. Also, the assumption on symmetry should be released; to start pushing and to stop pushing should be described with different rates. If these differences are introduced,
the equation of motion is
\begin{eqnarray*}
\label{eq10}
\frac{da}{dt}=a^2(-aw_{aa}+bu_{aa})+ab(-aw_{ab}+bu_{ab})+\\
ba(-aw_{ba}+bu_{ba})+b^2(-aw_{bb}+bu_{bb}),
\end{eqnarray*}
where $u_{xy}$ ($w_{xy}$) is the rate of start pushing (stop pushing) when a pedestrian behind applies the strategy $x$ and the pedestrian before applies $y$. When the equation \eqref{eq10} is rewritten as
\begin{eqnarray*}
\label{eq11}
\frac{da}{dt}=-a^3w_{aa}+b^3u_{bb}-a^2b(w_{ab}+w_{ba}-u_{aa})+\\
ab^2(u_{ab}+u_{ba}-w_{bb})
\end{eqnarray*}
the number of parameters falls to four. In a generic case, neither $m=0$, nor $m=\pm1$ is a solution. The stationary state can be found from the third-order algebraic equation by the Cardano method. As the number of real solutions can be either one or three, we expect that the related bifurcation is again of fork type.

As a special case, let us consider the case when there is no outflow from the states $aa$ and $bb$. Then, $w_{aa}=u_{bb}=0$,
and the equation of motion reduces to
\begin{equation}
\label{eq12}
\frac{dm}{dt}=2\frac{da}{dt}=\frac{1}{4}(1-m^2)(D-C-m(D+C)),
\end{equation}
where $C=w_{ab}+w_{ba}-u_{aa}$ and $D=u_{ab}+u_{ba}-w_{bb}$. There are three fixed points here, $m_{\pm}=\pm 1$ and $m_0=(D-C)/(D+C)$. If $D+C$ is positive, $m_0$ exists and is stable if both $C$ and $D$ are positive; having changed the sign of both $C$ and $D$, we have the same rule. In two remaining areas either $m_+$ or $m_-$ is stable; the respective conditions are $C <0$ and $D<0$. There is no more than one stable solution for a given set of parameters. The bifurcations are transcritical and not of fork type. As in the previous case, the linearization procedure allows to obtain the relaxation times analytically; for $m_+$ (stable if $C<0$) we get $\tau=-1/C$, for $m_-$ (stable if $D<0$) we get $\tau=-1/D$, and for $m_0$ (stable if $DC/(D+C)>0$) we get $\tau=(D+C)/(DC)$. In all cases, the relaxation time goes to infinity when the fixed point goes to the boundary of the area of its stability. Here these boundaries are $C=0$ and $D=0$. We expect this behaviour of $\tau$ also in more general cases.

The order-disorder phase transition in one-dimensional systems is usually an artifact of mean field theory. On the other hand, a natural medium to simulate processes in crowd is a two-dimensional lattice. As we have seen, for the case of two symmetric strategies we need five parameters $w_n$, $n=0,\cdots,4$. In the presence of an ordered phase the rate $w_n$ should decrease with $n$. This means, that there is a positive correlation between actions of neighboring pedestrians. Having this in mind, we have calculated numerically the magnetization on a square lattice, with the rates as follows: $w_0=1$, $w_1=2x$, $w_2=x$, $w_3=x/2$ and $w_4=x/4$. The choice of $w_0=1$ is equivalent to an assumption, that the strategy opposite to what all neighbours do is abandoned immediately. Other rates are set as that an increase of the number of neighbors who perform the same strategy by one leads to doubling the mean time when the strategy is continued. As we have no experimental data on these rates, the choice remains arbitrary. However, the decrease of $w_n$ with $n$ has some support in the results of observations of Mark Granovetter on the so-called critical mass effect\cite{threshold,sznajd}. The results of the simulation is shown in Fig. 1 as the curve of pluses. The approximate value of the transition point $x_C$ is 0.06. Note however, that for the latter red curve the parameter $x$ has no thermodynamic interpretation. The mean field equation \eqref{eq1} for these rates gives the stationary solution
\begin{equation}
\label{eq13}
m=a-b=\sqrt{\frac{4+5x-2\sqrt{16-56x+85x^2}}{15x-4}},
\end{equation}
with the critical point close to $x=0.571$. Here again, its value is overestimated by the mean field model.

%% ############################################################################
\section{Conclusions\label{sec4}}
%% ############################################################################

We have proposed a scheme of translation of thermal phase transitions to phase transitions related to an interplay of rates of different processes. This enables wide and easy application of results obtained in frames of statistical mechanics to other sciences, where temperature is not a handy concept. In particular, in social sciences we have plenty of examples when people change their social roles, professions, status etc.\cite{odmn}. The scheme provides a straightforward way to computer simulations, where the rates identify the probabilities of change of a single agent.

The scheme is applied to construct a model phase transition, where one of two phases is chosen: the selfish strategy and the cooperative strategy. These options are discussed in therms of a crowd or a queue, where pedestrians push the person before or stand still and move only when others move. We demonstrated, that at the points of the phase diagram where strategies are changed, the relaxation time of reaching the stationary phase goes to infinity.
This effect, known in statistical mechanics as the critical slowing down, is basically a consequence of an increase of spatial correlation at the critical point. It is known to be present in numerical simulations, but it is reproduced also in mean field solutions. If a measurement of the related human behaviour is possible, this effect could be detected.  

As we know, neither mean field theory nor computer simulations is a decisive method to state if a phase transition really exists. This is important in particular in the context of a one-dimensional Ising model with short range interaction. However, neither the thermodynamic limit nor the thermal equilibrium are met in social experiments, where systems are always finite and never stationary. What kind of conclusions can be drawn, then, from the mean-field one-dimensional model?  It seems reasonable to expect that if the states $m_{\pm}$ are close to be absorbing, they should appear in any finite system after a sufficiently long time. Then, the thermodynamic limit should leave them unchanged. The formal trick is therefore first to pass with time to infinity, later with the number of agents to infinity. However, even if the states $m_{\pm}$ are not fully absorbing, we should observe for appropriate areas in the parameter space long-lasting periods when the majority of pedestrians plays this or that strategy. The situation is somewhat similar to the persistence of the stripes in the ground state of a two-dimensional Ising ferromagnet, which can be annealed out in a very small temperature\cite{red}.

Concluding, all thermodynamic Monte Carlo simulations of magnetic phase transitions can be converted to athermal models, where the rates of switching of spins are treated as independent parameters. The rates are useful in Monte Carlo simulations, but they also enter to time-dependent mean field equations in a natural way. The athermal models can be used in areas other than physics of magnetism. For a sociologically motivated example of norm breaking in crowd, we have shown that the mean field approach reproduces the critical slowing down effect in the critical region.

%% ############################################################################
\section*{Acknowledgements} 
%% ############################################################################
One of authors (K.K) is grateful to Sidney Redner for his kind comment, and to Maria Nawojczyk for pointing us to reference\cite{odmn}. The research is partially supported within the FP7 project SOCIONICAL, No. 231288, and the Polish Ministry of Science and Higher Education.

\end{document}